\documentclass[aps,pra,twocolumn,superscriptaddress,longbibliography]{revtex4-2}

\usepackage{natbib}
\usepackage{pgfplots,amsfonts}
\usepackage[breaklinks]{hyperref}
\usepackage{amssymb}
\usepackage{amsmath,amsthm,bm}
\usepackage{amsfonts,dsfont}
\usepackage{graphicx}
\usepackage{lipsum}

\usepackage{subcaption}
\usepackage{mathtools}
\usepackage{tikz}
\usepackage{cleveref}
\usetikzlibrary{angles, quotes}
\usetikzlibrary{positioning}

\hypersetup{citecolor=red,colorlinks=true,urlcolor=blue}

\usepackage[normalem]{ulem}
\usepackage{color}
\usepackage{xcolor}

\newcommand{\bd}[1]{\boldsymbol{#1}}

\newcommand{\Tr}{\mbox{Tr}}

\newcommand{\bra}[1]{\mbox{$\langle #1 |$}}
\newcommand{\ket}[1]{\mbox{$| #1 \rangle$}}

\newcommand{\NCU}{Institute of Physics, Faculty of Physics, Astronomy and Informatics, Nicolaus Copernicus University in Toru\'n, Grudzi\c{a}dzka 5, 87-100 Toru\'n, Poland}
\newcommand{\BEC}{Pitaevskii BEC Center, CNR-INO and Dipartimento di Fisica, Università di Trento, I-38123 Trento, Italy}
\newcommand{\MPI}{Max Planck Institute for the Physics of Complex Systems,  N\"othnitzer Str.~38, 01187, Dresden, Germany}

\begin{document}

\title{
Extracting Many-Body Quantum Resources \\ 
within One-Body Reduced Density Matrix Functional Theory
}

\author{Carlos L. Benavides-Riveros}
\email{cl.benavidesriveros@unitn.it}\affiliation{\BEC} 
\affiliation{\MPI}

\author{Tomasz Wasak}
\email{twasak@umk.pl}\affiliation{\NCU}

\author{Alessio Recati}
\affiliation{\BEC}

\date{\today}

\begin{abstract}
Quantum Fisher information (QFI) is a central concept in quantum sciences used to quantify the ultimate precision limit of parameter estimation, detect quantum phase transitions, witness ge\-nuine multipartite entanglement, or probe nonlocality. Despite this widespread range of applications, computing the QFI value of quantum many-body systems is, in general, a very demanding task. Here we combine ideas from functional theories and quantum information to develop a novel functional framework for the QFI of fermionic and bosonic ground states. By relying upon the constrained-search approach, we demonstrate that the QFI matrix terms can universally be determined by the one-body reduced density matrix (1-RDM), avoiding thus the use of exponentially large wave functions. Furthermore, we show that QFI functionals can be determined from the universal 1-RDM functional by calculating its derivatives with respect to the coupling strengths, becoming thus the \textit{generating} functional of the QFI. We showcase our approach with the Bose-Hubbard model and present exact analytical and numerical QFI functionals. Our results provide the first connection between the one-body reduced density matrix functional theory and the quantum Fisher information.
\end{abstract}

\maketitle

\textit{Introduction.---} The concept of quantum correlations is transversal for many areas of quantum physics ranging from condensed matter \cite{Islam2015,Bergschneider2019} and quantum chemis\-try \cite{HUANG20051,C7CP01137G} to high-energy physics \cite{Calabrese_2009,Nishioka_2009}. Among different measures of quantum correlations, the quantum Fisher information (QFI) \cite{doi:10.1142/S0219749909004839,Amari2016} is a key quantity that not only gives an operational meaning for multipartite entanglement for quantum-enhanced metrology  of spins \cite{PhysRevLett.127.260501}, bosons~\cite{smerzi2009entanglement, toth2012multipartite,Wang_2014} and fermions~\cite{ hauke2021quench}, but can also be used to probe quantum criticality \cite{PhysRevB.96.104402, hauke2016measuring}, nonlocality \cite{PhysRevLett.109.020505,PhysRevLett.126.210506,PhysRevLett.120.040402}, and quantum geometry \cite{Lambert_2023}. This widespread range of applications makes QFI a fundamentally important concept in quantum phy\-sics \cite{Tóth_2014,Liu_2020,PhysRevLett.123.250502,10.1063/1.1697374}. However, due to the problem of finding an optimal measurement which yields the QFI,  its determination in quantum many-body systems remains an important theoretical and technological challenge \cite{RevModPhys.90.035005,PhysRevResearch.4.013083}.

Experimentally, the value of the QFI was extracted, in the form of a lower bound, and used to prove entanglement of non-Gaussian ma\-ny-body states~\cite{oberthaler2014nongaussian} and pair-correlated states of twin matter waves~\cite{klempt2011twin}. 
For thermal states, the QFI was directly related to dynamic susceptibilities~\cite{hauke2016measuring} and to the quench dynamics in linear response~\cite{hauke2021quench}.
With the former method, the QFI was measured and used to quantify entanglement in a spin-1/2 Heisenberg antiferromagnetic chain~\cite{nagar2020experimental} and for nitrogen-vacancy center in diamonds~\cite{yu2022quantum}.
Recently, a method to determine the optimal use of a given entangled state for quantum sensing was proposed based on the QFI matrix (QFIM)~\cite{holland2023generators} showing its usefulness to determine optimal quantum technology protocols. Despite this experimental progress, extracting quantum correlations in many-body systems, and, thus, quantifying their quantum resources through QFI, is still hampered by the Hilbert space's exponential growth, rendering the computation a formidably demanding task~\cite{PhysRevLett.120.240402,Batle2016}. 

A strategy to alleviate the cost of computing global quan\-ti\-ties of quantum many-body systems is to estimate them from local measurements. For instance, artificial neural networks can be trained to learn the entangle\-ment entropy or the two-point density correlations of interacting fermions from local correlations \cite{10.21468/SciPostPhysCore.6.2.030,PhysRevLett.125.076402,Shao2023,doi:10.1021/acs.jctc.2c00850,aikebaier2023machine}. The intuition behind this is that some physical properties of quantum systems (usually the ones of ground states) can unambiguously be determined by certain reduced quantities. Many rigorous theorems establish the existence of one-to-one maps between ground states $\ket{\psi}$ and appropriately chosen sets of reduced quantities $\{\varphi_\mu\}$, such as the particle density or the reduced density matrix, justifying thus the \textit{functional} notation $\ket{\psi[\{\varphi_\mu\}]}$~\cite{HK, UvonBarth_1972,10.1002/qua.20303,Gilbert,PhysRevB.99.224502}. As a consequence, all observables of the system's ground state are also implicitly functionals of those reduced quantities. Yet, while this is true, almost all research in functional theories focuses on developing energy functionals \cite{PhysRevLett.81.866,Jones15,PhysRevLett.127.233001,PhysRevLett.119.063002,bookciosolo,Pernal,Baerends,doi:10.1063/5.0139897,Cohen2,sharma,10.1063/5.0171981}. Questions about multipartite quantum correlations or nonlocality in the systems these functionals describe are usually neither addressed nor even posed \cite{PhysRevLett.128.013001}. 

Here we initiate and develop a functio\-nal-theo\-retical framework for the~QFI. We will show that for ground states of identical particles the QFIM can be determined by the one-body reduced density matrix (1-RDM) $\bm{\gamma}$, obtained by tracing out $N-1$ particles of the $N$-body quantum state, avoiding thus the pre-computation of wave functions that expand into exponentially large Hilbert spaces. This indicates a promising cross-fertilization of the theory of quantum resources with concepts and techniques developed in the functional theory of the 1-RDM (1-RDMFT), as it is currently used in quantum chemistry (where its fermionic version has been developed and, so far, mainly employed). We will unveil two surprising links between 1-RDMFT and the QFI functional theory introduced in this work: (i) \textit{the energy functional of the 1-RDM can be fully reconstructed from the functionals of the QFI} and (ii) \textit{QFI functionals correspond to the derivatives of the 1-RDM functionals with respect to the coupling strengths}, revealing thus the ability of 1-RDMFT to capture itself quantum correlations.

The paper is structured as follows: First, we present a general framework for the functional theory of the 1-RDM and recap the concept of QFIM. Next, we present the main ingredients of our novel functional theory of QFI, showcasing our approach for a Bose-Hubbard model. We finish with conclusions.  A Supplemental Material (SM) contains additional technical details.

\textit{Hubbard-like Hamiltonian.---} We begin with the Hamiltonian:
\begin{align}\label{H}
 \hat H = \hat h + \hat W.
\end{align}
The one-body part $\hat h = -\sum_{\langle ij\rangle} t_{ij} \hat b_i^\dagger \hat b_j$ describes the tunneling, while the two-body term is of the general form
\begin{align}\label{defW_general}
    \hat W = \frac12 \sum_{ijkl} V_{ijkl} 
        \hat b_{i}^\dagger \hat b_{j}^\dagger \hat b_{k}\hat b_{l}.
\end{align}
The operators $\hat b_j^\dagger$ ($\hat b_j$) create (annihilate) a particle on site~$j$. For a standard Hubbard model with on-site interaction, the couplings are nonzero only when all the indices are equal $V_{jjjj} = U$. We keep the general matrix elements to take into account non-standard Hubbard models accounting, for example, for dipolar couplings~\cite{dutta2015non}. For notational convenience and in order to relate the functional formalism to quantum information concepts, we define site-dependent angular-momentum Hermitian operators, i.e., $\hat J^{ij}_\alpha = \tfrac12 (\hat b_{i}^\dagger, \hat b_{j}^\dagger) \, \sigma_{\alpha} \, (\hat b_{i}, \hat b_{j})^T$, where $\sigma_\alpha$, with $\alpha=0,x,y,z$, are Pauli matrices. The two-body operator \eqref{defW_general} can be rewritten, up to a one-body operator 
that can be incorporated into $h$, as
\begin{align}\label{Hubbard}
    \hat W = \sum_{\alpha, \beta} \sum_{ijkl} u_{\alpha\beta}^{ijkl} \{\hat J_\alpha^{ij},\hat J_\beta^{kl}\},
\end{align}
where $\{A,B\} = AB+BA$ and $\bm{u} = \{u^{ijkl}_{\alpha\beta}\}$ are real coupling strengths subject to additional constraints stemming from the symmetry of the operators $\{\hat J_\alpha^{ij},\hat J_\beta^{kl}\}$.  

One of our results is that the universal interacting functional $\mathcal F[\bm{\gamma};\bm{u}]$ that appears in the framework of 1-RDMFT for describing the ground-state expectation value of $\hat W$ \eqref{Hubbard} serves as a generating functional for the functionals of QFI matrix elements. We first introduce both concepts ($\mathcal F$ and QFI) and then show this connection.

\textit{One-body reduced-density-matrix functional theory.---}
Given a $N$-body density matrix $\rho$, we define the 1-RDM as
\begin{align}
    \gamma_{\alpha}^{ij} \equiv \mathrm{Tr}\big[\rho \hat J_{\alpha}^{ij}\big].
\end{align}
Here, ($\alpha$, $i$, $j$) is the collective index of the vector $\bm\gamma$. 
Although this representation of $\bd\gamma$ is not standard \cite{Pernal2016}, it will be convenient to develop our functional formulation of QFI. The ground-state problem for a many-body Hamiltonian of the form in Eq.~\eqref{H} can be solved without resorting to wave functions by proving the existence of a 1-RDM-functional~$\mathcal{F}[\bm{\gamma};\bm{u}]$ \cite{Gilbert}, which describes the two-body interactions in terms of $\bm\gamma$: for any choice of $h$ the ground-state 1-RDM follows from the minimization of  $\mathcal{E}[\bm{\gamma}] = \Tr[h \bm{\gamma}] + \mathcal{F}[\bm{\gamma};\bm{u}]$, where the first term depends linearly on $\bm\gamma$. Since the functional $\mathcal{F}$ depends only on the \textit{fixed} interaction $W$ (not on the one-particle Hamiltonians) it is called \textit{universal}~\footnote{Eventually, the functional depends also on the chosen (fer\-mio\-nic/bosonic) statistics of the problem and the total number of particles. Consequently, for a specific problem fixed by $\hat h$, the relevant 1-RDM fulfills the equation $\nabla_{\bm{\gamma}} \mathcal{F}[\bm{\gamma}] = - h$ resulting from the minimization of~$\mathcal E[\bm\gamma]$}. The \textit{constrained-search approach} \cite{Levy6062} indicates a route for the calculation of $\mathcal{F}$ by minimizing the expectation value of the interacting energy over all states $\ket\psi$ that lead to the same $\bm{\gamma}$. Symbolically,
\begin{align}
\mathcal{F}[\bm{\gamma};\bm{u}] &= \min_{\psi \rightarrow \bm{\gamma}} \bra{\psi}\hat W \ket{\psi} \nonumber \\
&=  
    \min_{\psi \rightarrow \bm{\gamma}} 
    \sum_{\alpha, \beta} \sum_{ijkl} u_{\alpha\beta}^{ijkl}
     \bra{\psi} \{\hat J_\alpha^{ij},\hat J_\beta^{kl}\} \ket{\psi}.
     \label{CSA}
\end{align}
So far, a few explicit (exact) calculations of this functional ha\-ve been carried out for bosonic \cite{PhysRevLett.124.180603,PhysRevResearch.3.013282,Maciazek_2021,PhysRevResearch.3.L032063} and fer\-mionic \cite{Schilling2019, Cohen1} systems. Interestingly, the corresponding mi\-ni\-mi\-zers  $\ket{\psi[\bm{\gamma};\bm{u}]} = {\rm argmin}_{\psi \rightarrow \bm{\gamma}} \bra{\psi}  W \ket{\psi}$ do not only correspond to ground states but to the entire set of (representable) 1-RDMs \footnote{In the literature of functional theory the minimizers of the constrained-search are called $v$-representable when they correspond to the ground state of some Hamiltonian (with the same $W$) or non-$v$-representable when they not. The concept of $N$-re\-pre\-sen\-tability refers to 1-RDMs that come from at least one $N$-particle quantum state. }. Let us highlight that for $W$ in Eq.~\eqref{Hubbard}, the functional \eqref{CSA} depends explicitly both on the 1-RDM and on the entire set of coupling constants $\bd{u}$. This is a known fact in 1-RDMFT, recently emphasized by Cioslowski and coauthors \cite{Cioslowski2,Cioslowski1} for the case of the functional using the repulsive term \eqref{defW_general}. Below we show how, when taking \eqref{Hubbard} instead of \eqref{defW_general},  the QFI can be directly extracted from $\mathcal{F}$ when the latter is considered as a generating functional for both repulsive and attractive interactions.

\textit{Quantum Fisher information.---} Consider a transformation of the density matrix $\rho$, describing the state of the quantum system, using a unitary operator~$\hat U(\bm\phi)$. The resulting state becomes $\rho(\bm\phi) = \hat U(\bm\phi) \rho \hat U^\dagger(\bm\phi)$. The distance between these two density matrices (and the response of the quantum state to perturbations) can be quantified by the QFI. Specifically, for the response $\partial \rho(\bm{\phi}) / \partial \phi_a \equiv (\rho \hat L_a + \hat L_a \rho)/2$, where $\hat L_a$ is called the Symmetric Logarithmic Derivative and $\phi_a$ is the $a$-th parameter of the vector $\bm{\phi}$, the QFIM is $M_{ab} = \mathrm{Tr}[\rho \hat L_a \hat L_b]$.
Since large QFI implies high sensitivity, originally the QFI was introduced in the context of quantum metrology and sensing via the quantum Cramér-Rao lower bound, which quantifies the ultimate 
precision limit in estimation protocols~\cite{Braunstein1994statistical, holevo2011probabilistic, RevModPhys.89.035002}. 

We parametrize the unitary with the angular-mo\-men\-tum operators, i.e., $\hat U(\bm\phi) = \exp(i \sum_{\alpha} \sum_{ij} \phi_{\alpha}^{ij} \hat J_{\alpha}^{ij})$. For pure states, which are considered in this work, $\rho = \ket{\psi}\bra{\psi}$, the QFIM becomes the covariance matrix  
\begin{align}
    M_{\alpha\beta}^{ijkl}[\psi] = 2\bra{\psi}\{\hat J_\alpha^{ij}, \hat J_\beta^{kl} \}\ket{\psi} - 4\bra{\psi}\hat J_\alpha^{ij}\ket{\psi} \bra{\psi}\hat J_\beta^{kl} \ket{\psi}.
    \label{Fisherpure}
\end{align}
An impediment to the calculation of this matrix is the prior knowledge of the quantum state, which involves computations of such complexity that hampers broad applicability. This problem, we will now see, can be circumvented by relying upon the 1-RDMs and the universal functional $\mathcal{F}$. 

We note that the QFI is an entanglement measure~\cite{smerzi2009entanglement, toth2012multipartite}: for a quantum system of $N$ bosons or spins, the state exhibits at least $(m+1)$-particle entanglement~\cite{hyllus2012fisher} if the single-parameter QFI surpasses the quantum limit, e.g., $\sum_{\alpha\beta} n_\alpha n_\beta M_{\alpha\beta}^{ijij} > s m^2 + (N-sm)^2$, where $s$ is an integer part of $N/m$, and $m= 1,2,3,\ldots$ quantifies entanglement depth; the vector $n$ is normalized, $\sum_\alpha n_\alpha^2=1$. For $m=1$, the right-hand side, equal to $N$, is the standard quantum limit~\cite{RevModPhys.90.035005}. Similar inequality was derived for fermions~\cite{hauke2021quench}. If the inequality is violated, the entanglement structure is not revealed.

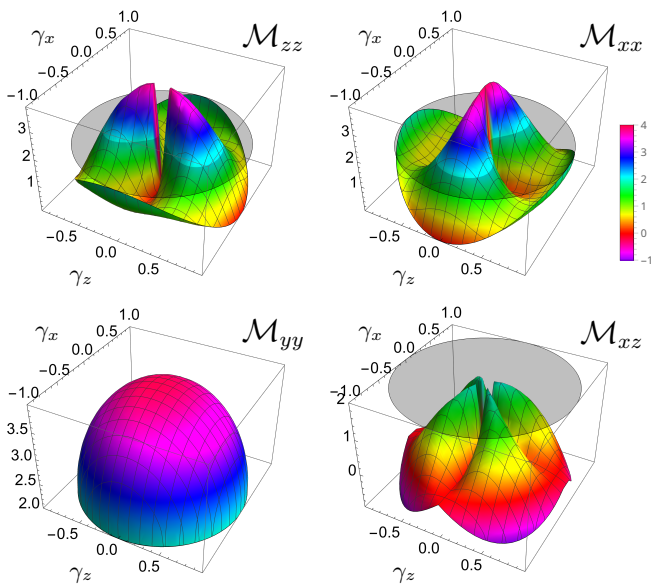
\begin{figure}[t]
\begin{tikzpicture}
 \node (img) {
  \includegraphics[scale=0.29]{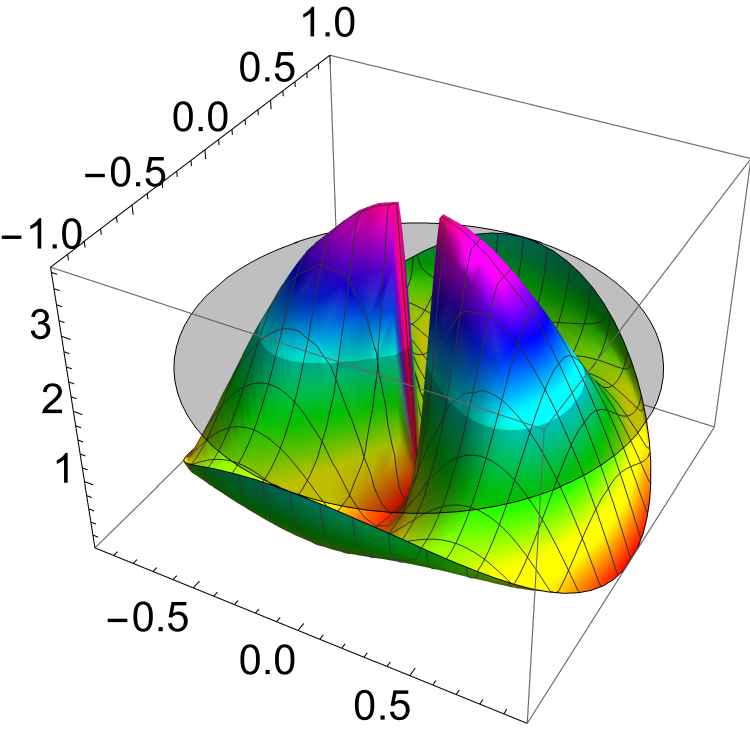}\hspace{0.5cm}
  \includegraphics[scale=0.29]{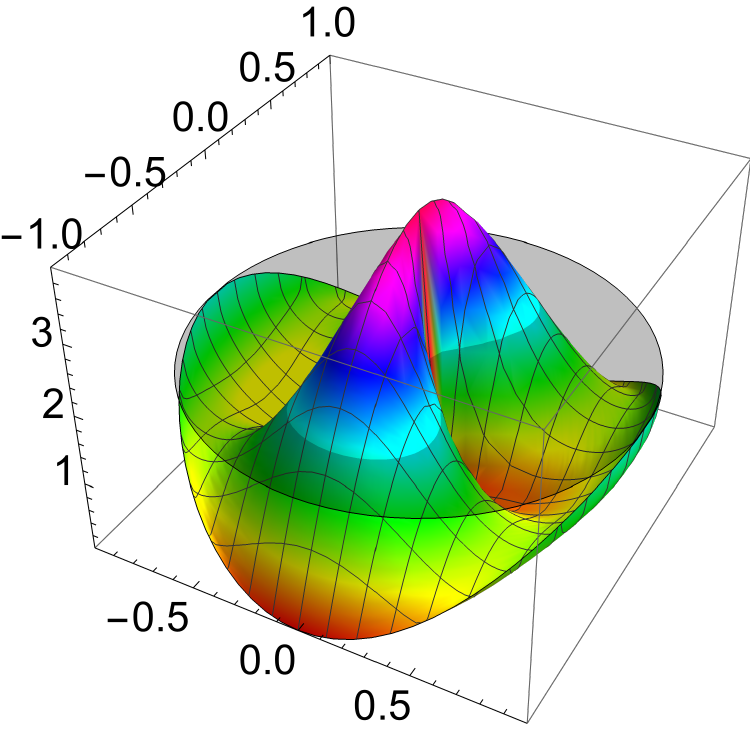} \includegraphics[scale=0.4]{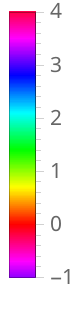}};
    \node[left=of img, node distance=0cm, anchor=center, xshift=2.1cm,yshift=-1.8cm,font=\color{black}] {$\gamma_{z}$};
    \node[left=of img, node distance=0cm, anchor=center, xshift=6.4cm,yshift=-1.8cm,font=\color{black}] {$\gamma_{z}$};
  \node[left=of img, node distance=0cm, anchor=center, xshift=1.6cm,yshift=1.4cm,font=\color{black}] {$\gamma_{x}$};
  \node[left=of img, node distance=0cm, anchor=center, xshift=5.95cm,yshift=1.4cm,font=\color{black}] {$\gamma_{x}$};
    \node[left=of img, node distance=0cm, anchor=center, xshift=4.65cm,yshift=1.4cm,font=\color{black}] {\large $\mathcal{M}_{zz}$};
       \node[left=of img, node distance=0cm, anchor=center, xshift=9.15cm,yshift=1.4cm,font=\color{black}] {\large $\mathcal{M}_{xx}$};
 \end{tikzpicture}
 \begin{tikzpicture}
 \node (img) {
  \includegraphics[scale=0.29]{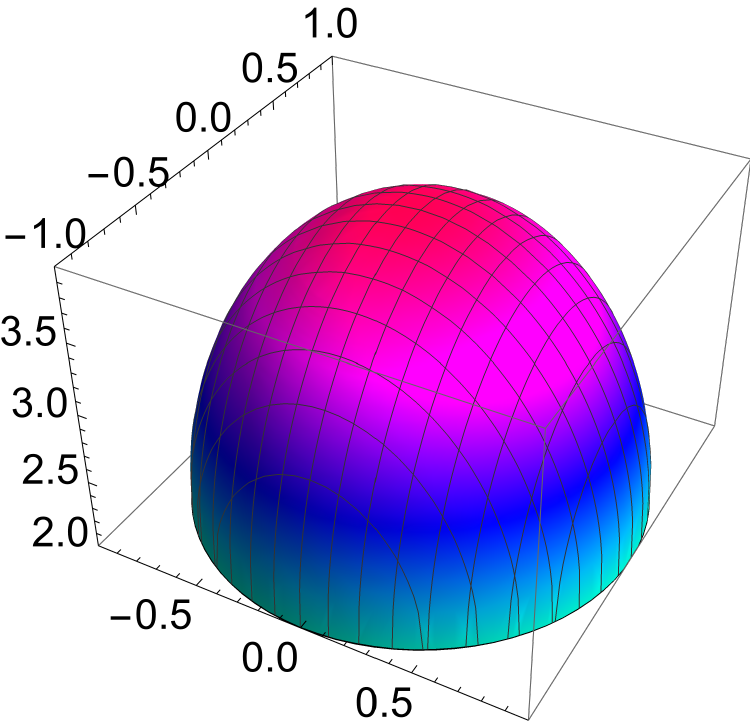}\hspace{0.5cm}
  \includegraphics[scale=0.29]{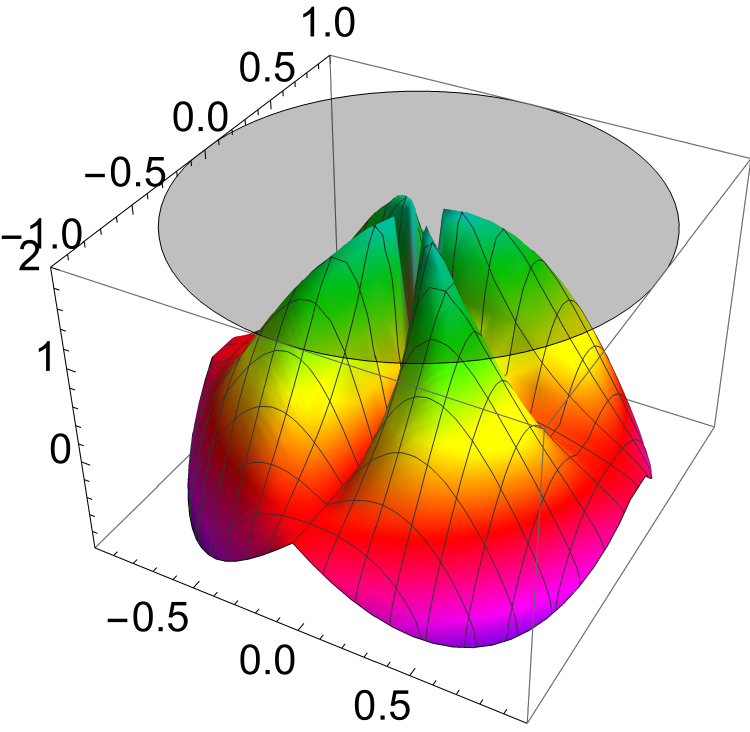}};
    \node[left=of img, node distance=0cm, anchor=center, xshift=2.1cm,yshift=-1.8cm,font=\color{black}] {$\gamma_{z}$};
  \node[left=of img, node distance=0cm, anchor=center, xshift=5.95cm,yshift=1.4cm,font=\color{black}] {$\gamma_{x}$};
  \node[left=of img, node distance=0cm, anchor=center, xshift=1.65cm,yshift=1.4cm,font=\color{black}] {$\gamma_{x}$};
    \node[left=of img, node distance=0cm, anchor=center, xshift=6.4cm,yshift=-1.8cm,font=\color{black}] {$\gamma_{z}$};
    \node[left=of img, node distance=0cm, anchor=center, xshift=4.65cm,yshift=1.4cm,font=\color{black}] {\large $\mathcal{M}_{yy}$};
    \node[left=of img, node distance=0cm, anchor=center, xshift=9.15cm,yshift=1.4cm,font=\color{black}] {\large $\mathcal{M}_{xz}$};
 \end{tikzpicture}
\caption{Universal functionals of QFIM for the repulsive Bose-Hubbard model (for all $U>0$) for $N=2$. The limit $\mathcal{M}_{\alpha\beta} = 2$ is indicated as a disk in gray.}
\label{fig1}
\end{figure}

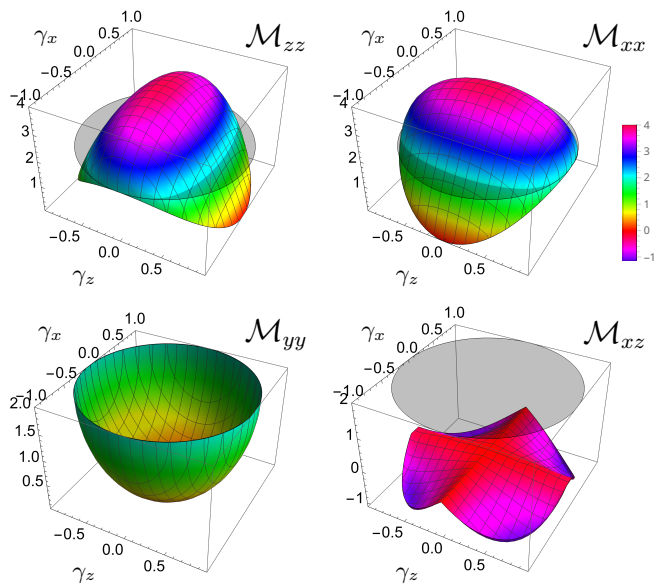
\begin{figure}[t]
\begin{tikzpicture}
 \node (img) {
  \includegraphics[scale=0.29]{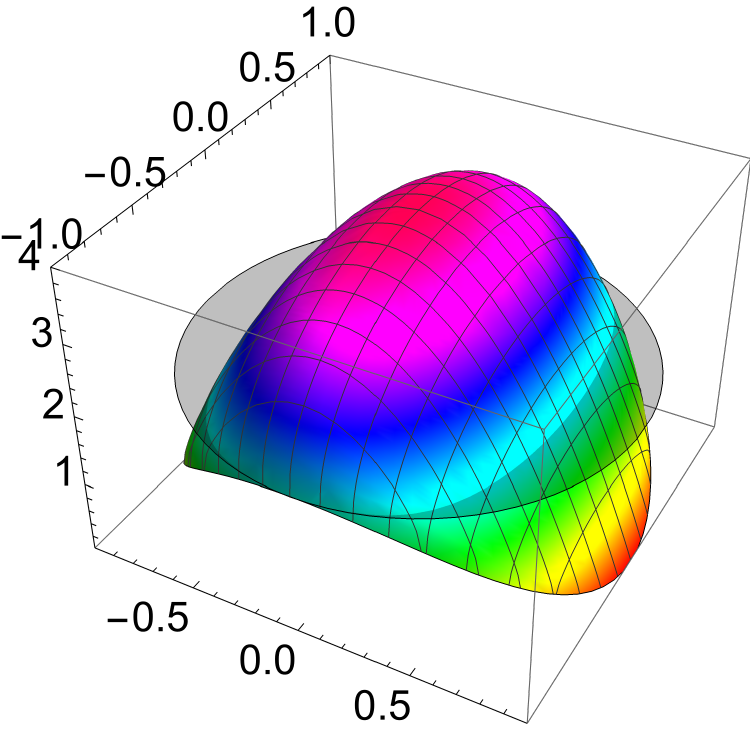}\hspace{0.5cm}
  \includegraphics[scale=0.29]{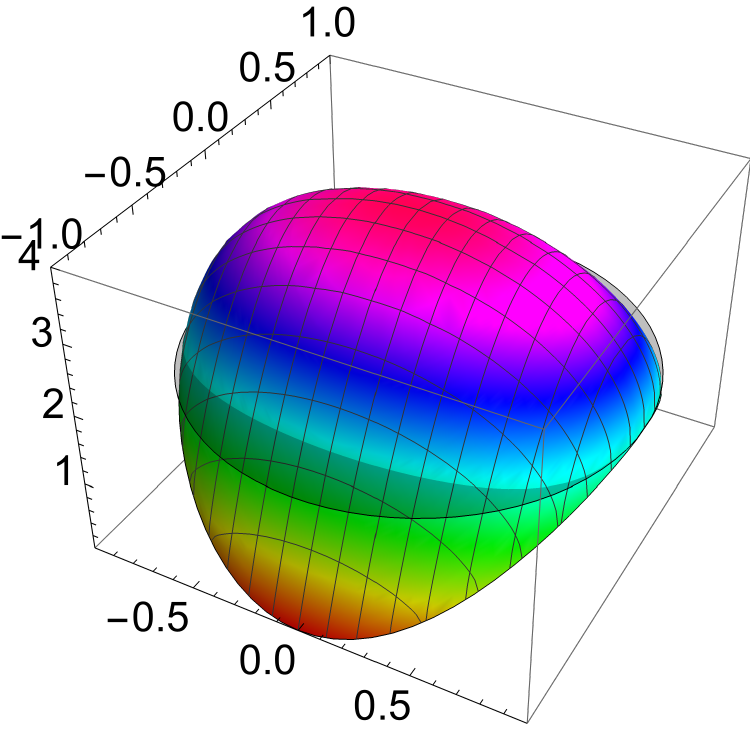} \includegraphics[scale=0.4]{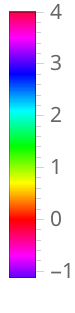}};
    \node[left=of img, node distance=0cm, anchor=center, xshift=2.1cm,yshift=-1.8cm,font=\color{black}] {$\gamma_{z}$};
    \node[left=of img, node distance=0cm, anchor=center, xshift=6.4cm,yshift=-1.8cm,font=\color{black}] {$\gamma_{z}$};
  \node[left=of img, node distance=0cm, anchor=center, xshift=1.6cm,yshift=1.4cm,font=\color{black}] {$\gamma_{x}$};
  \node[left=of img, node distance=0cm, anchor=center, xshift=5.95cm,yshift=1.4cm,font=\color{black}] {$\gamma_{x}$};
    \node[left=of img, node distance=0cm, anchor=center, xshift=4.65cm,yshift=1.4cm,font=\color{black}] {\large $\mathcal{M}_{zz}$};
       \node[left=of img, node distance=0cm, anchor=center, xshift=9.15cm,yshift=1.4cm,font=\color{black}] {\large $\mathcal{M}_{xx}$};
 \end{tikzpicture}
 \begin{tikzpicture}
 \node (img) {
  \includegraphics[scale=0.29]{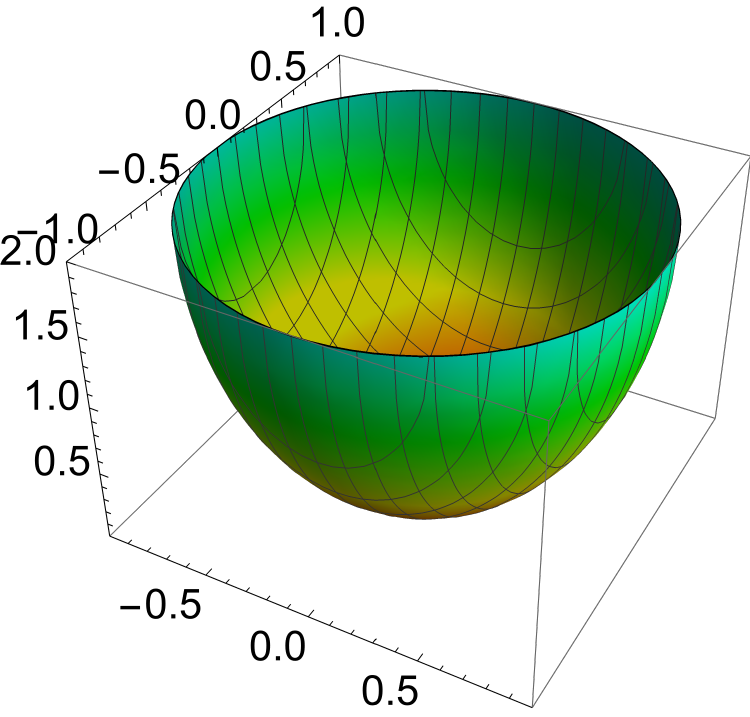}\hspace{0.5cm}
  \includegraphics[scale=0.29]{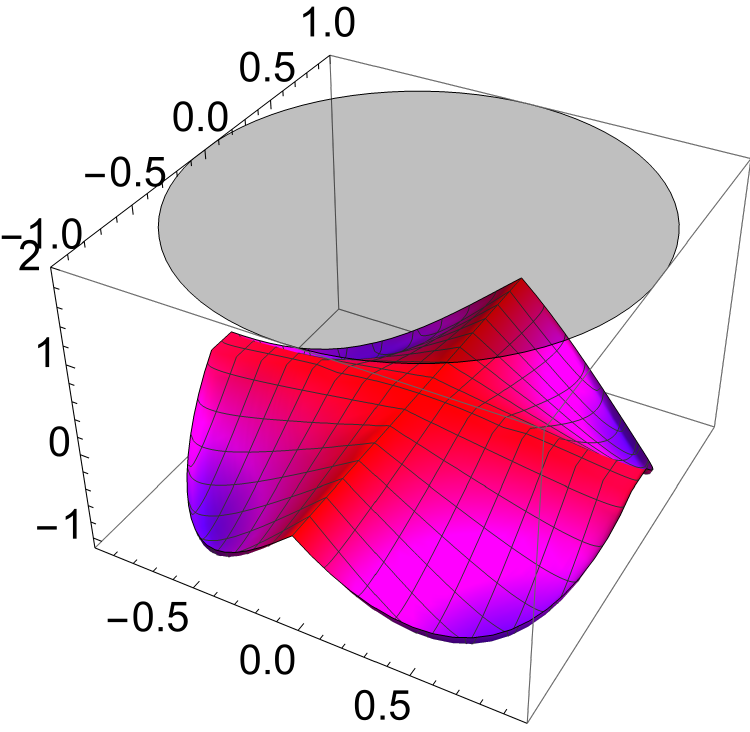}};
    \node[left=of img, node distance=0cm, anchor=center, xshift=2.1cm,yshift=-1.8cm,font=\color{black}] {$\gamma_{z}$};
  \node[left=of img, node distance=0cm, anchor=center, xshift=5.95cm,yshift=1.4cm,font=\color{black}] {$\gamma_{x}$};
  \node[left=of img, node distance=0cm, anchor=center, xshift=1.65cm,yshift=1.4cm,font=\color{black}] {$\gamma_{x}$};
    \node[left=of img, node distance=0cm, anchor=center, xshift=6.4cm,yshift=-1.8cm,font=\color{black}] {$\gamma_{z}$};
    \node[left=of img, node distance=0cm, anchor=center, xshift=4.65cm,yshift=1.4cm,font=\color{black}] {\large $\mathcal{M}_{yy}$};
    \node[left=of img, node distance=0cm, anchor=center, xshift=9.15cm,yshift=1.4cm,font=\color{black}] {\large $\mathcal{M}_{xz}$};
 \end{tikzpicture}
\caption{Universal functionals of QFIM for the attractive Bose-Hubbard model (for all $U<0$) for $N=2$. The limit $\mathcal{M}_{\alpha\beta} = 2$ is indicated as a disk in gray.}
\label{fig2}
\end{figure}

\textit{Functional theory of QFIM.---} The central quantity of the framework presented above is the 1-RDM. Noteworthy, it already contains information about quantum correlations, via the correlation entropy $S[\bm{\gamma}] = -\Tr[\bm{\gamma} \ln \bm{\gamma}]$ \cite{10.21468/SciPostPhysCore.6.2.030,Gersodrf,Tichy_2011,Benavides-Riveros_2014}. Surprisingly, 1-RDMFT is mainly focused on the goal of computing the ground-state energy and some associated observables~\cite{Pernal2016,PhysRevA.107.022210,GR19,10.21468/SciPostChem.1.2.004,doi:10.1063/1.5109009,D1CP01742J,10.21468/SciPostPhys.14.5.120}, but this powerful formalism has not been used to scrutiny multipartite entanglement or nonlocality. Indeed, the map $\bm{\gamma},\bd{u} \to \ket{\psi[\bm{\gamma},\bd{u}]}$ can be used for the calculation of functionals of QFIM: by using Eq.~\eqref{Fisherpure} one can view QFIM as explicitly, universal functionals dependent on the 1-RDM:
\begin{align}
\mathcal{M}_{\alpha\beta}^{ijkl}[\bm{\gamma},\bd{u}] \equiv  M_{\alpha\beta}^{ijkl}[\psi[\bm{\gamma},\bd{u}]] .
\end{align}
Let's emphasize that this functional is defined in a domain whose degrees of freedom do not scale with the num\-ber of particles. This is our central result. It points out that QFIM can be defined in the domain of 1-RDMs and shows that the relevant degrees of freedom of the problem scale with the dimension of the 1-particle Hilbert space. No reference to the quantum state $\ket{\psi[\bm{\gamma}]}$ is needed. We now explain how $\mathcal{M}_{\alpha\beta}^{ijkl}$ can be directly obtained from the universal functional $\mathcal{F}$.

\textit{Generation of QFI and reconstruction  of $\mathcal F$.---} 
The universal functional is a function of the couplings $\bm{u}$, cf. Eq.~\eqref{CSA}. By applying the Hellmann-Feynman theorem to the derivative of the functional~\cite{PhysRevA.59.3359}, we show, as detailed in SM, that $\mathcal{M}$ can be generated by the total energy functional $\mathcal{E}$ by means of the relation
\begin{align} \label{derF}
    \mathcal{M}_{\alpha\beta}^{ijkl}[\bm{\gamma};\bd{u}] = 4\left[
    \left(\frac{\partial \mathcal{E}[\bm{\gamma};\bd{u}]}{\partial u_{\alpha\beta}^{ijkl}}\right)_{\!\!\bm{\gamma}}
    -\gamma_\alpha^{ij} \gamma_\beta^{kl} \right],
\end{align}
where $\bm\gamma$ is kept fixed in the derivative and we indicated that $\mathcal{M}$ depends on $\bd{u}$.
We note that one can substitute here $\mathcal F$ instead of $\mathcal{E}$, but the latter is accessible experimentally.

It is worth emphasizing that, despite being closely related to the commonly-known two-electron density cumulants that arise from the differentiation of the functional using Eq.~\eqref{defW_general}, the QFI matrix elements obtained from  Eq.~\eqref{derF} are very different in their physical content. For instance, the real-valued QFI matrix elements $\mathcal{M}^{ijkl}_{\alpha\beta}$, equaling certain linear combinations of the 2-cumulants (which, in general, are complex numbers; see Eq.~(7) of Ref.~\cite{Cioslowski1}), are measurable. They play, in fact, a fundamentally important role in quantum metrology with a rich physical interpretation that, in contrast, is only possible for certain traces and contractions of the 2-cu\-mu\-lant.

Quite remarkably, many results of 1-RDMFT straightforwardly apply to our framework. Firstly, as a result of the Hellmann-Feynman theorem,  Eq.~\eqref{derF} shows that the (QFI generating) functional $\mathcal{F}[\bd{\gamma},\bd{u}]$ is positive homogeneous of degree 1 in the coupling constants $\bd{u}$ \cite{Cioslowski1,Cioslowski2,10.1063/1.1516804}. Secondly, this implies that the functionals of QFI enter (after adding a correcting product of 1-RDMs) directly in the universal functional:
\begin{align}
    \mathcal{F}[\bm{\gamma};\bd{u}] = 
        \sum_{\alpha\beta}\sum_{ijkl}
        u_{\alpha,\beta}^{ijkl}\left(\frac14 
        \mathcal{M}_{\alpha,\beta}^{ijkl}[\bm{\gamma};\bd{u}] + 
        \gamma_\alpha^{ij}\gamma_\beta^{kl}\right).
\end{align}
Hence, the QFIM, together with its content about quantum resources, enters into the universal functional. These results show that the knowledge of the QFIM allows for full reconstruction of the universal 1-RDM functional. Finally, mirroring an analogous property of the 2-cumulant \cite{Cioslowski1}, as a result of the homogeneity of the functional $\mathcal{F}$ it follows from Euler's theorem that QFIM is zero-degree positive-homogeneous in the coupling constants: $\sum_{\alpha\beta;ijkl}
u_{\alpha,\beta}^{ijkl}\partial\mathcal{M}_{\alpha,\beta}^{ijkl}[\bm{\gamma};\bd{u}]/\partial u_{\alpha,\beta}^{ijkl} = 0$. 
We note that this does not imply the QFIM is independent of $\bd{u}$. As shown below for the simplest Hubbard model with a single coupling strength $u$, the QFIM depends on the sign of $u$.


\textit{Two-well Bose-Hubbard model.---}  
Two-well Hubbard models played a historical role as analytical tests for density functional theory in its ground-state \cite{Carrascal2015}, time-dependent \cite{Carrascal2018}, and excited-state ensemble \cite{10.1063/1.5084312} versions in order to unveil analytical properties of the functionals \cite{Cohen1,PhysRevLett.124.180603,10.1063/5.0143657,Be23}. It was already used in the context of bosonic Josephson Junctions to obtain quantum resources in terms of spin squeezing and QFI \cite{oberthaler2014nongaussian,Alicki_2023}. An advantage of the bosonic model is that, while it can be filled with an arbitrary number of particles, the functionals can be visualized as 3D graphs. The Hamiltonian is:
\begin{align}
    \hat H = -t\sum_{\langle ij \rangle} \hat b_i^\dagger \hat b_j  + u\sum_{j} \hat n_j (\hat n_j -1),
\end{align}
where $\hat n_j= \hat b^\dagger_j \hat b_j$ is the particle-number operator in the site $j=r$ (right) or $l$ (left); the on-site interactions in Eq.~\eqref{defW_general} are $V_{rrrr} = V_{llll} = u$ and the rest are zero. The term $\hat W =  u\sum_{j \in\{l,r\}} \hat n_j (\hat n_j -1)$ is the relevant quantity of what follows. As only two sites are considered we drop the site indices:
\begin{align}
\label{gamma_J}
    \gamma_\alpha = \bra{\psi}\hat J_\alpha \ket{\psi},
\end{align}
with $\sum_\alpha \gamma_\alpha^2 \leqslant N^2/4$. This means that all 1-RDMs  lie inside the Bloch sphere of radius $N/2$.

The universal functional contains all the information to reconstruct one of the diagonals of the QFIM. This is a consequence of the relation between $\hat W$ and  $\hat J_z$:
\begin{align}
\sum_{j \in \{l,r\}} \hat n_j (\hat n_j -1) =  2\hat J_z^2 +\frac{\hat N^2}{2}-\hat N,
\end{align}
where $\hat N = \hat n_l + \hat n_r$. 
Hence, since $\ket{\psi}$ has a fixed number of particles, we obtain: 
$\bra{\psi}\hat W \ket{\psi}/u = 2 \bra{\psi}\hat J_z^2 \ket{\psi} +N^2/2-N$. By replacing $\psi \rightarrow \psi[\bm{\gamma}]$, it follows
\begin{align}
    \mathcal{M}_{zz}[\bm{\gamma}] = 4\left(\frac{\mathcal{F}[\bm{\gamma};u]}{u}  -\gamma_z^2 \right) -N^2+2N,
\end{align}
which is a special case of Eq.~\eqref{derF} with a single coupling strength.
It is instructive to write the expression of the functional for $N=2$. With real wave functions ($\gamma_y = 0$), 
\begin{align}
    \mathcal{M}_{zz}[\bm{\gamma}] = 4\left[1-\frac{1+\sqrt{1-(\gamma_x^2+\gamma_z^2)}}{2(\gamma_x^2+\gamma_z^2)}\gamma_x^2-\gamma_z^2\right].
\end{align}

In Fig.~\ref{fig1} the exact functionals for $\mathcal{M}_{xx}$, $\mathcal{M}_{yy}$, $\mathcal{M}_{zz}$ and $\mathcal{M}_{xz}$ are presented for $U>0$. Since the amplitudes of the wavefunctions are real and $\{\hat J_x, \hat J_y \}$ and   $\{\hat J_y, \hat J_z \}$ are skew symmetrical operators, $\mathcal{M}_{xy} = \mathcal{M}_{yz} = 0$,  everywhere.
With a gray disk, we mark the value of the standard quantum limit and thus the values $\mathcal{M}_{\alpha\alpha}>N=2$ signal entanglement. We observe that for $\mathcal{M}_{yy}$, all the states are entangled apart from the surface of the Bloch sphere, describing spin coherent states, at which the quantum limit is not surpassed also for $\alpha=x,z$, i.e., $\mathcal{M}_{\alpha\alpha} \leqslant N$.

We notice that the above results are valid for repulsive interaction, and, therefore, not all quantum states can become a minimizer. For instance, the NOON state $(\ket{2,0}+\ket{0,2})/\sqrt{2}$ shows up in the functionals with attractive interactions $U<0$. They are sketched in Fig.~\ref{fig2}. While some features are similar, they differ greatly from the functionals in Fig.~\ref{fig1}: for the attractive ca\-se, the value of the QFI of most of the ground states lies above the quantum limit, although there are states \textit{within} the Bloch sphere which do not exhibit entanglement.

\begin{figure}[!t]
\begin{tikzpicture}
 \node (img) {
  \hspace{2.175cm}\includegraphics[scale=0.565]{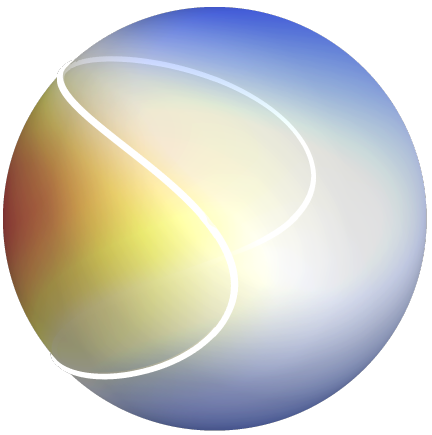}\hspace{1.4cm}\includegraphics[scale=0.45]{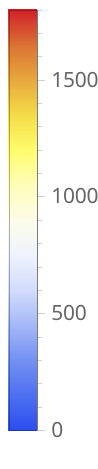}};
  \def\r{2}  
  \draw (0,0) node[circle, fill, inner sep=1] (orig) {} -- (-\r/3-0.3,\r/2)
    node[circle, fill, color=red,inner sep=2,label=above:{\Huge \color{black}$\bm{\gamma}$}] (a) {};
  \draw[dashed] (orig) -- (-\r/3, -\r/8)  node[circle, fill, color=black,inner sep=1,label=below right:{}] (phi) {};
  \draw[dashed] (orig) ellipse (\r{} and \r/3-0.1);
  \draw[->] (orig) -- ++(\r/3-0.3, \r/3-0.1) node[above] (x2){\color{black} \Large $\bm\gamma_{y}$};
  \draw[->] (orig) -- ++(\r, 0) node[right] (x1) {\color{black} \Large $\bm\gamma_{x}$};
  \draw[->] (orig) -- ++(0, \r) node[above] (x3) {\color{black} \Large $\bm\gamma_{z}$};
  \pic [draw=black, text=black, <->, "$\varphi$",angle eccentricity=0.6] {angle = phi--orig--x1};
  \pic [draw=black, text=black, <->, "$\theta$", angle eccentricity=1.4] {angle = x3--orig--a};
    \draw (0,0) node[circle, fill, inner sep=1] (orig) {} -- (-\r/3-0.3,\r/2)
    node[circle, fill, color=red,inner sep=2,label=above:{\Huge \color{black}$\bm{\gamma}$}] (b) {};
\end{tikzpicture}
\caption{Representation of $\bm{\gamma}$, parametrized with the angles $\theta$ and $\varphi$, inside the Bloch sphere of radius $N/2$. The color-code represents the value of $\mathcal{M}_{zz}$ close to BEC for $N=1000$ and $\delta = 0.1$ (see Eq.~\eqref{bec-M}). The white line marks the standard quantum limit $\mathcal M_{zz}=N$. 
}
\label{Blochfig}
\end{figure}
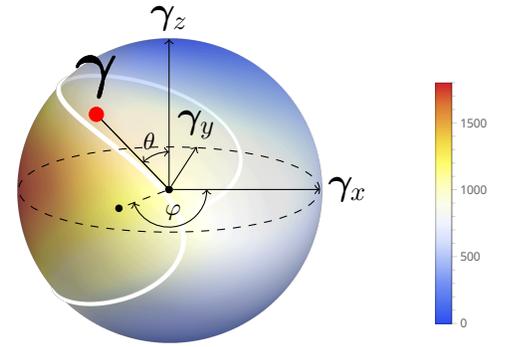

\textit{QFIM functional theory for BEC.---} Another interesting limit of the two-well Bose-Hubbard model is the Bose-Eins\-tein condensate (BEC) state, which has been used for quantum metrological tasks, spin squeezing and test Bell correlations \cite{Schmied441,Fadel409,MA201189,PhysRevA.93.033617}. Due to the Penrose-Onsager criterion \cite{Penrose1956}, BEC states lie near the border of the Bloch sphere and it is convenient to define the functionals in terms of the angles $\theta$ and $\varphi$, and the number of particles depleted from the condensate (i.e., away from the set of spin coherent states): $\delta = N/2 - \gamma$ (see Fig.~\ref{Blochfig}). Thus, the QFIM functionals can be expressed as functions of the spherical coordinates $\mathcal{M}_{\alpha \beta}(\delta, \theta,\varphi)$. In the neighborhood of the condensation point,  $\mathcal{M}_{zz}$ reads (see SM):
\begin{align}
    &\mathcal{M}_{zz}(\delta, \theta, \varphi)
    = \mathcal M_{zz}^{(0)}(\theta)
    - \mathcal M^{(1/2)}_{zz}(\theta,\varphi) {\delta^{1/2}} \nonumber\\ 
    & \qquad \qquad \qquad  \qquad 
    + \mathcal M_{zz}^{(1)}(\theta) \delta 
    + \mathcal{O}(\delta^{3/2}).
    \label{bec-M}
\end{align}
Here, $\mathcal{M}_{zz}^{(0)}(\theta) = N \sin^2(\theta)$ represents the mean-field fluctuations which cannot surpass  the standard quantum limit, i.e., $\mathcal{M}_{zz}^{(0)}(\theta) \leqslant N$, 
whereas the two beyond-mean field corrections scaling as $\delta^{1/2}$ and $\delta$ are $\mathcal{M}^{(1/2)}_{zz}(\varphi,\theta) = 2\sin^2(\theta)\cos(\varphi) \sqrt{N(N -1)}$ and $\mathcal{M}^{(1)}_{zz}(\theta) = 8 + 2 (N - 6) \sin^2(\theta)$, respectively. The square-root scaling of the second term in Eq.~\eqref{bec-M} yields the diverging BEC force, that drives nonperturbatively the system away from the mean-field state \cite{PhysRevLett.124.180603,PhysRevResearch.3.013282,Maciazek_2021}. Since the first term does not violate the standard quantum limit, only the next orders can contribute to genuine multipartite entanglement. In Fig.~\ref{Blochfig} an example of $\mathcal{M}_{zz}(\delta, \theta,\varphi)$ is shown for $N=1000$ and for depletion $\delta = 0.1$. To surpass the standard quantum limit, the mean field contribution should be large, and thus $\theta\approx \pi/2$, so $\bm{\gamma}$ lies close to the equator. Next, the third term in Eq.~\eqref{bec-M} is always positive, but is significantly overshadowed by the second term in the region $\cos\varphi>0$. However, in the region $\cos\varphi<0$ it contributes positively, and we observe enhancement of entanglement.

\textit{Conclusion.---} In this paper we combined ideas from functional theories and quantum information to develop a functional approach to the QFI. In our formalism the elements of the QFI matrix are \textit{functionals} of the 1-RDM, avoiding thus the exponential growth of the Hilbert space in which they are usually defined. We obtained two main results: (i) \textit{the knowledge of the QFIM allows the full reconstruction of the universal functional of the 1-RDM} and (ii) \textit{QFI functionals correspond to the derivatives of the 1-RDM functional with respect to the coupling strengths}, the latter being upgraded to the level of generating functional of the QFIM functionals. These results show a so far unexplored ability of the 1-RDM functionals to detect genuine multipartite entanglement. Since in 1-RDM functional theory approach, we can freely adjust single-particle Hamiltonians to move in the landscape of the functionals, our work shows a novel way to extract many-body resources and to determine optimal sensing protocols~\cite{holland2023generators}.
\\

The data presented in this article is available from~\cite{data}.

\begin{acknowledgments}
We acknowledge the European Union’s Horizon Europe Re\-search and Innovation program  un\-der the Marie Skło\-dowska-Curie GA n°101065295-RDMFTforbosons. This research is part of project No. 2021/43/P/ST2/02911 co-funded by the National Science Centre and the European Union’s Horizon 2020 Re\-search and Innovation program under the Marie Skłodowska-Curie GA  n°945339.
For the purpose of Open Access, the author has applied a CC-BY public copyright licence to any Author Accepted Manuscript (AAM) version arising from this submission. This work has been funded from Provincia Autonoma di Trento. 
\end{acknowledgments}

\bibliography{Refs2}

\end{document}